\documentclass[twocolumn,showpacs]{revtex4-1}
\usepackage{latexsym}
\usepackage{graphicx}
\usepackage{amssymb}
\usepackage{amsmath}
\usepackage{epstopdf}
\usepackage{color}

\definecolor{azu}{rgb}{0.0,0.4,1.0}

\usepackage{ulem}

\begin{document}
\title{The influence of antiphase boundary of the MnAl $\tau$-phase on the energy product}
\author{S. Arapan$^{1,2}$}
\email{sergiu.arapan@gmail.com}
\author{P. Nieves$^{2}$}
\author{S. Cuesta-L\'opez$^{3}$} 
\author{M. Gusenbauer$^4$}
\author{H. Oezelt$^4$}
\author{T. Schrefl$^4$}
\author{E. K. Delczeg-Czirjak$^5$}
\author{H. C. Herper $^5$}
\author{O. Eriksson$^{5,6}$}

\affiliation{$^1$ IT4Innovations, VSB-Technical University of Ostrava, 17. listopadu 15, CZ-70833 Ostrava-Poruba, Czech Republic}
\affiliation{$^2$ ICCRAM, International Research Center in Critical Raw Materials and Advanced Industrial Technologies, University of Burgos, 09001 Burgos, Spain}
\affiliation{$^3$ ICAMCyL, International Center for Advanced Materials and Raw Materials of Castilla y L\'eon, 24492 L\'eon, Spain}
\affiliation{$^4$ Department for Integrated Sensor Systems, Danube University Krems, 2700 Wiener Neustadt, Austria}
\affiliation{$^5$ Department of Physics and Astronomy, Division of Materials Theory, Uppsala University, Box 516, SE-75120 Uppsala, Sweden}
\affiliation{$^6$ School of Science and Engineering, \"{O}rebro University, SE-701 82 \"{O}rebro, Sweden}
\date{\today}                                         

\begin{abstract}
In this work we use a multiscale approach toward a realistic design of a permanent magnet based on  MnAl $\tau$-phase  and elucidate how the antiphase boundary defects present in this material affect the energy product. We show how the extrinsic properties of a microstructure depend on the intrinsic properties of a structure with defects by performing micromagnetic simulations. For an accurate estimation of the energy product of a realistic permanent magnet based on the MnAl $\tau$-phase with antiphase boundaries, we quantify for the first time the exchange interaction strength across the antiphase boundary defect with a simple approach derived from the first-principles calculations. These two types of calculations performed at different scales are linked via atomistic spin dynamic simulations performed at an intermediate scale. 
\end{abstract}
\pagebreak

\maketitle

\begin{figure}[h!] 
   \centering
    \includegraphics[width=\columnwidth ,angle=0]{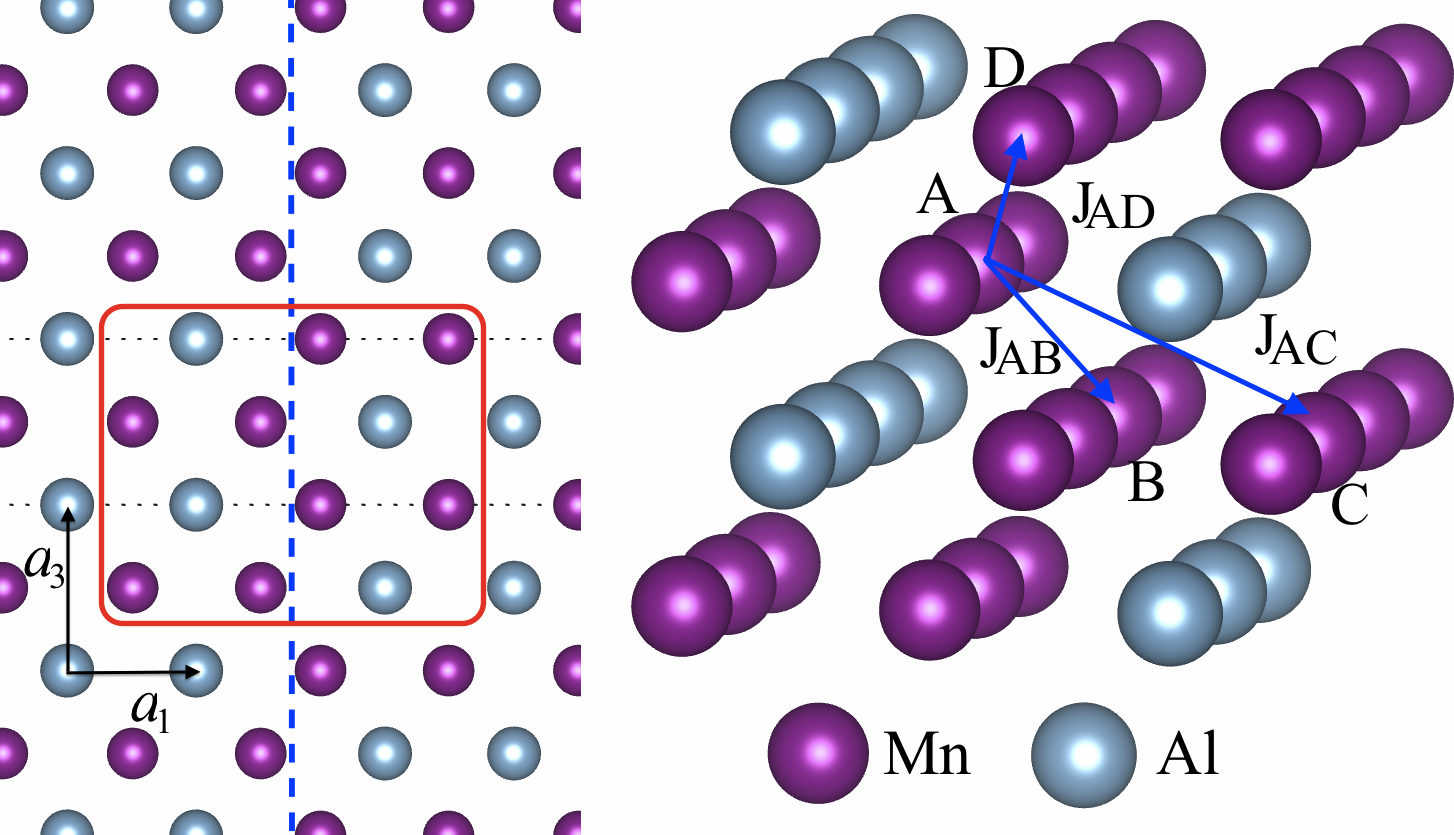} 
   \caption{ (left) An ideal APB created by a shear of the L1$_0$ lattice by $(\vec{a}_{1}+\vec{a}_{3})/2$ in the [101] direction, where $|\vec{a}_{1}|$ and $|\vec{a}_{3}|$ are the lattice parameters of the L1$_0$ cell. The vertical dashed line shows the interface between two domains with collinear spin configurations: either magnetic moments on Mn ions point in the same (a FM arrangement) direction or in opposite directions (an AFM arrangement) in different regions. Thin dotted lines show the boundary of the $N\times1\times1$ supercell used in calculations. (right) A close up at the local environment of a Mn ion at the interface (the region enclosed by the rectangle on the left hand side of the figure). For the Mn ion A at one side of the APB interface the first three closest Mn neighbors across the interface, B, C, and D are shown.}
   \label{fig:APB}
\end{figure}

MnAl has attracted a considerable interest since the discovery of the ferromagnetic $\tau$-MnAl phase in the early 1960s~\cite{Kono,Koch}. The phase is metastable and forms for a concentration of Mn ions slightly higher than 50\%. Mn and Al ions occupy the sites of the ordered L1$_0$ structure. This ferromagnetic phase exhibits a modest magnetization of 0.60 MA/m and an energy product $(BH)_{max}$ of 50 kJ/m$^3$~\cite{Coey}, but is appealing due to abundance and low cost of needed raw materials. However, the discovery of Nd-Fe-B permanent magnets had suspended further improvements of the $\tau$-MnAl ferromagnets for several decades. Although the advance in the energy product of rare-earth (RE) based magnets are not making further progress, it is their critical supply status that calls for developing alternative permanent magnets. From this perspective the $\tau$-MnAl phase with an estimated upper limit of $(BH)_{max}\approx$120~kJ/m$^3$ could fill the niche between high-performance RE permanent magnets and general purpose ferrites. In practice, the maximum energy product of a magnetic material is affected by its microstructure, which, in turn, is determined by the methods of preparation. The L1$_0$ ordered MnAl shows potential in sense, and exhibits various types of defects, like anti-phase boundaries (APBs)~\cite{Zijlstra} and twins~\cite{Bittner}. After the formation of the $\tau$-phase the coercivity of the material is relatively low because of the presence of APBs~\cite{Houseman}. Both types of defects, APBs and twins may act as pinning centers for domain walls~\cite{Jakubovics,Thielsch}. APBs may also be responsible for easy nucleation of domain walls. They could be created by a shear of the L1$_0$ lattice in the [101] direction by $(\vec{a}_{1}+\vec{a}_{3})/2$ giving rise to an anti-ferromagnetic (AFM) coupling between nearest Mn ions (Fig.\ref{fig:APB}). As a result, the crystal is made of two antiparallel ferromagnetic regions, which lowers the energy product the system can host~\cite{Zijlstra}. Recent micromagnetic simulations have confirmed that both types of defects, APBs and twins, facilitate the domain wall nucleation and pinning~\cite{Bance}. The granular structure containing fine twining and APBs may reduce the $(BH)_{max}$ to just 5\% of its theoretical maximum and should be avoided during the manufacturing process. 

In a recent paper~\cite{Pnieves}, we have performed atomistic spin dynamics (ASD) simulations~\cite{Eriksson} on two-phase region of the  MnAl $\tau$-phase to analyze the magnetic properties at the APB. Within the ASD the interacting magnetic ions are approximated by an effective 3D Heisenberg spin system and the dynamics is described by the stochastic Landau-Lifshitz-Gilbert (LLG) equations. By using the exchange coupling at the interface as a free parameter we have observed in the strong coupling regime the reorientation in the hard plane of the local magnetic moments of Mn atoms at the interface. This suggests that the nucleation process is favored at the APB if a strong antiferromagnetic interaction between Mn atoms exists at the interface. The critical value of the exchange interaction between Mn atoms at the APB, which leads to the formation of an APB, decorated with a domain wall, was calculated to be $\sim -10$ meV~\cite{Pnieves}.

In this work we aim to elucidate how the APB defects, decorated with a domain wall, influence the energy product of a permanent magnet based on MnAl $\tau$-phase. To obtain sensible results, we determine from ab-initio calculations the antiferromagnetic interaction between Mn atoms at the APB. We show that there indeed exists a strong antiferromagnetic interaction between Mn ions across the APB, which favors the nucleation process of reversed domains at very low field values at the defect. Using the obtained value of the antiferromagnetic exchange interaction between Mn atoms across the APB, we estimate the room temperature values of the anisotropy constant, saturation magnetization, and exchange stiffness via ASD calculations, and use them as an input to the micromagnetic simulations of realistic microstructure. We show in detail how the density of APBs lowers the coercive filed and decreases the energy product.    

Determining the antiferromagnetic interaction between Mn atoms at the APB is central to this work. Usually, the exchange integrals $J_{ij}$ used to describe the interaction of spins in the Heisenberg model, are estimated from ab-initio calculations performed for the bulk. $J_{ij}$ are calculated generally by one of two methods. In one method one calculates the $J_{ij}$ via the Korringa-Kohn-Rostoker (KKR)-Green function formalism by considering the rotation of two spin moments at sites $i$ and $j$ with opposite angles and calculating the total energy variation~\cite{Liechtenstein}. This method readily provide the magnetic exchange interaction within the muffin-tin or atomic sphere approximation for metals, as well as for binary alloys through KKR-CPA scheme~\cite{Faulkner}. Another method is to use the frozen-magnon approach by calculating the total energy for a spiral magnetic configuration~\cite{Rosengaard}. Both methods are formally equivalent and complementary to each other~\cite{Pajda}. Within the framework of frozen-magnon approximation, the magnetic configuration is constrained to a spin-spiral with the wave vector $\mathbf q$ and the spin-wave energy $E(\mathbf q)$ is calculated. Then, the exchange parameters $J_{ij}$ are obtained by Fourier-transformation. 

In principle, one could advance both methods to calculate the exchange parameters for a structure with planar defects like interfaces or sandwiched layers with different magnetic properties. Here, however, we propose a basic approach to determine the $J_{ij}$ for atoms located at the interface, which is independent of the implementation of ab-initio method used to calculate the total energy of a magnetic structure. In the case of the APB it is based on the observation that the energy of magnetic interaction at the interface, $E_{\text{APB}}$, between two magnetic domains is mainly determined by the interaction between neighboring Mn ions at the interface. By interface energy per atom we understand the energy of a Mn atom at the interface in the average field created by surrounding atoms. The interface energy has mainly two contributions: one coming from spin-spin interaction within the same domain, and second from spin-spin interaction across the APB. The first contribution is identical for both collinear ferromagnetic (FM) and antiferromagnetic (AFM) spin configuration across the APB (see Ref. \cite{Pnieves}). The AFM configuration corresponds to a domain wall with a width equal to one lattice spacing.  Let us denote the interaction between the Mn atoms located at positions $(0,0,0)$ and their first nearest neighbors at $(1/2,1/2,1/2)$ by $J_{AB}$, and between their second nearest neighbors by $J_{AC}$ and $J_{AD}$ (these second nearest neighbors are located at slightly different separation distances due to the relaxation of atomic  positions at the APB). Then the second contribution to the interface energy per atom is given by  $\pm 4 (J_{AB}+J_{AC}+J_{AD}+...)$  for structures with AFM and FM collinear arrangements of spins across the planar interface, respectively (Fig.\ref{fig:APB}). The second argument is that the energy difference between structures with AFM and FM collinear spin configurations across the interface is given by the difference of the interface energies, $\Delta E=E_{\text{APB}}^{\text{AFM}}-E_{\text{APB}}^{\text{FM}}$. The exchange parameters decrease with the interatomic distance. The second nearest  Mn neighbors across the interface are located at distances almost twice larger than separation between nearest Mn atoms, thus, we may assume that the main contribution to the interface energy comes from the $J_{AB}$ term. In such a case, we can relate the difference in the interface energies of two configurations with the exchange integral between two neighboring Mn atoms: $J_{AB}=\Delta E/8$.  In the case that the second nearest neighbors interactions are not negligible, then the $J_{AB}$ calculated here should be understood as an effective exchange interaction across the APB consistent with the coarse-grained model developed in Ref.~\cite{Pnieves}. Now, the problem of determining the $J_{ij}$ parameter is reduced to calculating the interface energy (the energy of a Mn atom at the planar defect). 
One may object that determining the interface energy is still not a trivial problem. We have realized that we can solve this problem by performing supercell calculations. In practical calculations the system with a single planar defect in the (100) plane is modeled as a supercell comprising two domains of $\tau$-MnAl periodically repeated in the [100] direction (Fig.\ref{fig:APB}). The energy of the system with two regions of given collinear spin configuration is obtained in the limit of an infinitely large supercell. It is obvious, that the energy of the supercell, $E_{SC}$, divided by the number of atoms in the supercell, $N_{SC}$, should, in the limit of large supercell,  converge to the value of the energy per atom in the bulk $\tau$-phase MnAl, $E_{bulk}$. The total energy of a supercell can be factorized into two terms, one accounting for the energy of atoms from the bulk like region, $E_{bulk} N_{bulk}$, plus the energy of atoms from the interface, $E_{\text{APB}} N_{\text{APB}}$, where $N_{bulk}+N_{\text{APB}}=N_{SC}$. Usually, a supercell comprises an integral number of unit cells. Due to the periodic boundary conditions  our supercell model of the planar defect has two interfaces, thus two unit cells correspond to the interface region and the dependence of the energy per atom of a supercell as a function of the number of unit cells, $N_{cell}$, can be parametrized (according to the Heisenberg model) as:
\begin{eqnarray}
E(N_{cell}) & = & \frac{E_{bulk} N_{bulk} +E_{\text{APB}} N_{\text{APB}}}{N_{SC}}\nonumber\\
            & = & \frac{a(N_{cell}-2) +2b}{N_{cell}},
\label{eq:1}
\end{eqnarray}
where parameters $a$ and $b$ would correspond to the $E_{bulk}$  and $E_{\text{APB}}$, respectively. Performing a set of energy calculations for different sizes of the supercell we can fit energy values to the Eq.~\ref{eq:1} and obtain the energy of a Mn atom at the interface $E_{\text{APB}}=b$.          

\begin{figure}[h!] 
   \centering
    \includegraphics[width=\columnwidth ,angle=0]{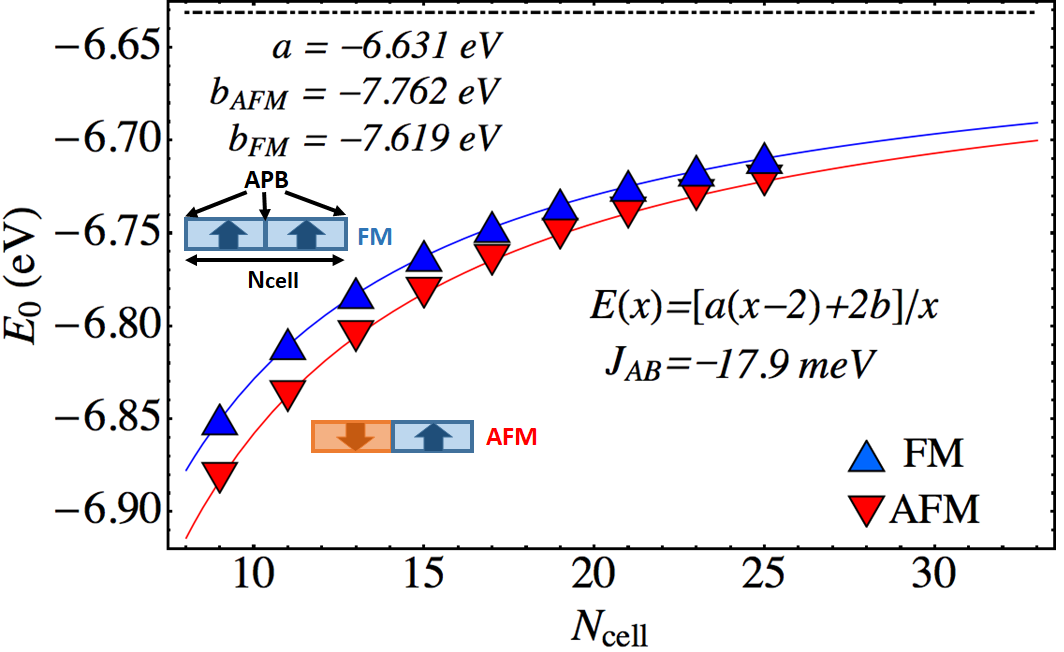} 
   \caption{ Ab-initio energy per atom vs the size of supercells modeling APB with FM (triangle up) and AFM (triangle down) collinear spin arrangements in the two domains across the boundary. Solid lines show the energy dependence $E=E(x)$, where $x=N_{cell}$, described by Eq.~\ref{eq:1} with parameters $a$ and $b$ fitted from calculations. The dash-dotted line shows the $\displaystyle\lim_{x\to\infty} E(x)$, which is the energy/atom of bulk MnAl L1$_{0}$ phase. }
   \label{fig:E0}
\end{figure}

We have modeled the $\tau$-MnAl structure with ideal APB as a set of supercells of sizes $2N-1\times1\times1$, where $N=5,6,...,13$. Total energy of the system was calculated by using the VASP code~\cite{VASP1,VASP2}. We have used PAW PBE 5.4 potentials~\cite{PAW} and an equally dense k-point mesh in the Brillouin zone. The energy cut-off was set to 450 eV and ions and supercell shape were optimized at each calculation. Performing a structural optimization is important for an accurate description of the interaction across the APB. With increasing the supercell size the average lattice parameters tend towards the $\tau$-MnAl bulk values, but the rate of convergence is quite slow. The analysis of interatomic distances shows that about 1\% difference compared to the bulk structure occurs mainly in the vicinity of the APB, while atoms beyond next-nearest neighbors from the interface are bulk-like spaced even for small sizes of supercell. Finally, the estimation of the relaxation energy shows a similar (about 2 meV/atom difference) contribution to the interface energy due to lattice optimization for both types of APB. We show results of supercell calculations in Fig.~\ref{fig:E0}. We can see that Eq.~\ref{eq:1} perfectly fits calculated energy values and gives for the parameter $a=-6.631$ eV exactly the ground state energy for the bulk L1$_0$ MnAl~\cite{Pnieves} structure for both, FM and AFM collinear spin arrangements across the interface. This fitting provides us also with the energy  $b$ of the Mn atoms at the interface, and the difference between these energies for two different types of APB is $\Delta b=b_{\text{AFM}}-b_{\text{FM}}= -143$ meV/atom. This difference comes entirely from different collinear spin arrangements across the interface and allow us to determine the value of the exchange integral to be $J_{AB}=\Delta b/8=-17.9$ meV. 

 Our calculated value of the exchange interaction between Mn atoms at the APB is almost two times stronger than the critical one~\cite{Pnieves}. Thus, we demonstrate that the antiferromagnetic interaction between Mn atoms at the interface of a MnAl $\tau$-phase with APBs is strong enough to form domain walls at these defects, as it is observed in experiments~\cite{Houseman,Jakubovics,Jaku,Land}. Additionally, the fact that the energy per atom in supercells with APBs is lower than in bulk ($a$), see Fig. \ref{fig:E0}, suggests that the formation of these planar defects may be energetically favorable and explains why they frequently form and stabilize in the synthesis processes.

\begin{figure}[h!] 
	\centering
	\includegraphics[width=0.5\columnwidth ,angle=0]{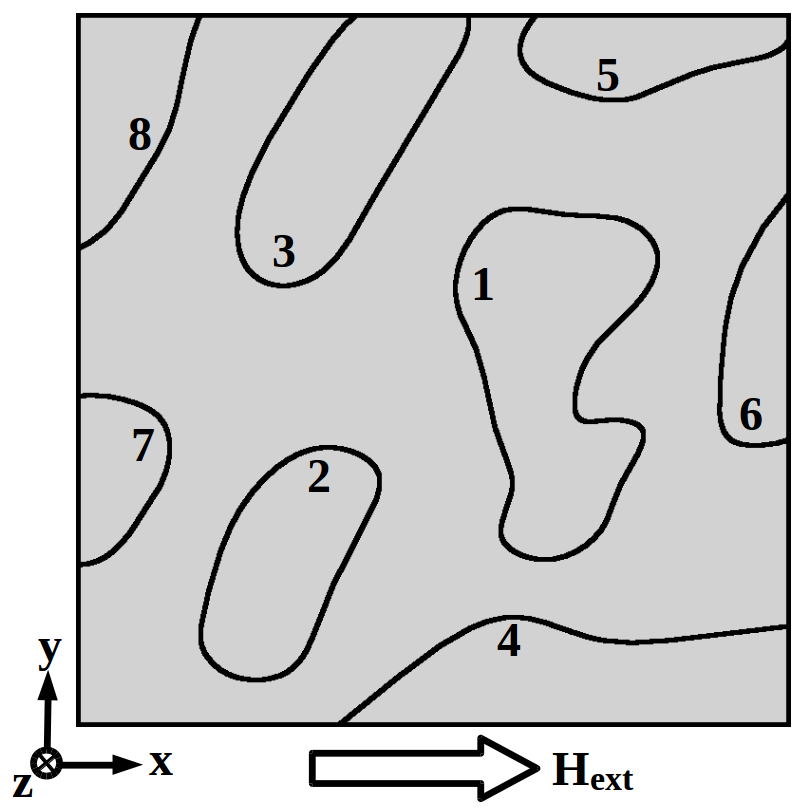} 
	\caption{Micromagnetic model of APBs. The geometry is obtained by a detail from a TEM image by Yanar et al.~\cite{Yanar} with edge lengths of 500 nm and a depth of 50 nm. The APBs are numbered consecutively and can be activated in their respective order. The external field is applied in-plane.}
	\label{fig:M0}
\end{figure}

The presence of APBs has a significant influence on the magnetization reversal process. APBs act as nucleation sites for reversed domains and as pinning sites for domain walls. An antiphase boundary helps to initiate magnetization reversal. However, once a reversed domain is formed the domain wall gets pinned at the crystallographic defect. Using the room temperature material properties of MnAl (anisotropy constant $K = 0.7$~MJ/m$^3$, a magnetization $\mu_0$M$_{\mathrm s} = 0.8$~T and an exchange stiffness $A = 7.6$~pJ/m) derived from first principle and ASD simulations \cite{Pnieves}, we computed the influence of APBs on magnetization reversal by micromagnetic simulations. We created a micromagnetic model from a TEM micrograph of APBs published in 2001~\cite{Yanar} (Fig.~\ref{fig:M0}). In the model we neglected other defects such as twins~\cite{Bittner}. An APB is modeled by an antiferromagnetic coupling across the interface, whereby we use the same model as for antiferromagnetic coupling through thin Ru layers in magnetic recording media~\cite{Dittrich,Bance}. Assuming a collinear AFM spin configuration across the APB and taking the exchange integral at the interface $J_{AB}=-17.9$~meV we derive an interface exchange energy density of $4J_{AB}/(a_1\cdot a_3)=-120$~mJ/m$^2$, where $a_1=0.276$ nm and $a_3=0.347$ nm are the lattice parameters of MnAl $\tau$-phase. At room temperature, taking into account a reduction of the micromagnetic exchange at the interface by thermal activation of both sides of the APB ($\propto m_e(T)^2$, where $m_e(T)$ is the equlibrium reduced magnetization at temperature T of bulk MnAl obtained by ASD~\cite{Pnieves}) we arrive at an effective interface exchange energy density of $-71$~mJ/m$^2$ (note that a higher value is obtained by rescaling temperature to experimental $m_e(T)$ in the ASD model~\cite{Pnieves,Evans}). 

\begin{figure}[h!] 
	\centering
	\includegraphics[width=\columnwidth ,angle=0]{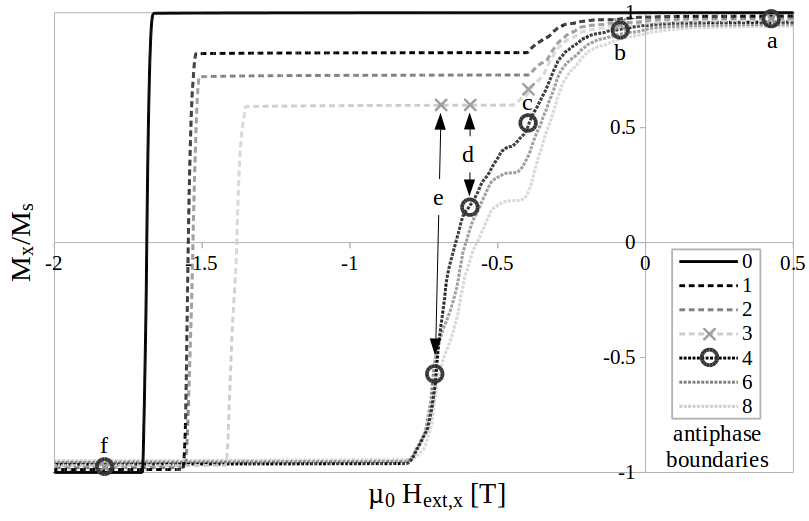} 
	\caption{Micromagnetically computed demagnetization curves of a grain containing eight APBs. The external field is applied in-plane. The amount of APBs activated is increased in their respective order. Letters a to f indicate magnetic states shown in Fig.~\ref{fig:M2}.} 
	\label{fig:M1}
\end{figure}

Fig.~\ref{fig:M1} shows the demagnetization curves for 0 to 8 activated APBs of the model in Fig.~\ref{fig:M0} using the above mentioned parameters. The APBs are activated consecutively in their respective order. The remanent magnetization decreases with increasing number of antiphase boundaries per grain.  The reason why demagnetization curve in Fig. \ref{fig:M0} has an unusual shape it is because we simulate a single grain, including more grains gives a typical, smooth and symmetric M vs H curve. The mechanism which explains the decrease of coercive field with the number of defects is the following: defects in the center of the model act as pinning sites (1--3 APBs), however defects close to the boundary nucleate the sample at low fields (4--8 APBs). Similarly, the energy density product decreases with increasing number of ABPs. The energy density product decreases rapidly, ranging from $(BH)_{\mathrm max} = 114$~kJ/m$^3$ for 1 APB per grain to $(BH)_{\mathrm max} = 84$~kJ/m$^3$ for 8 APBs per grain. Typically, in experiment one observes a $(BH)_{\mathrm max}$ around 50 kJ/m$^3$~\cite{Coey}, which suggests that there are more APBs per grain in the experimental samples. This is indeed correct, because we have chosen only a small section of the TEM image in ~\cite{Yanar}. Present calculations can reproduce observations with the number of APB being a determining factor. 

\begin{figure}[h!] 
	\centering
	\includegraphics[width=\columnwidth ,angle=0]{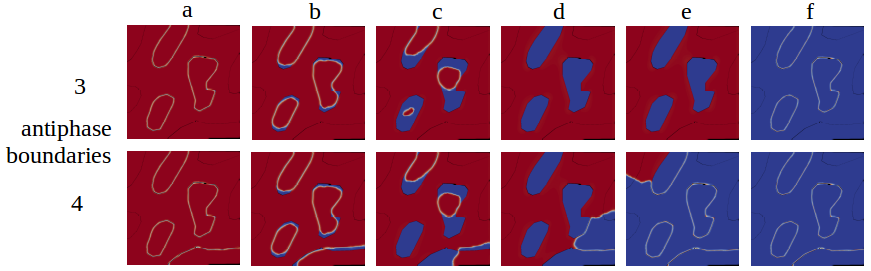} 
	\caption{Magnetic states of 3 and 4 activated APBs. Centered defects act as pinning sites (1--3 APBs) wheres APBs close to the boundaries nucleate the material already at low field values (4--8 APBs).}
	\label{fig:M2}
\end{figure}

In summary, we have performed numerical simulations at different scales to estimate the energy product of a realistic permanent magnet based on MnAl $\tau$-phase with APBs. First, we have obtained for the first time the value of the antiferromagnetic exchange interaction between Mn atoms across the APB by performing supercell electronic structure calculations. The calculated exchange interaction at the APB of MnAl $\tau$-phase is strong enough to form an APB decorated with a domain wall, which increase the phase stability, explaining why it is hard to avoid the formation of these defects in the synthesis processes. Second, we used previously calculated room temperature values of the anisotropy constant, saturation magnetization, and exchange stiffness by ASD simulations as input for micromagnetic simulations of the model of a MnAl $\tau$-phase with APBs. The micromagnetic modeling shows that ABPs deteriorate the loop shape through nucleation of reversed domains at very low field values and successive domain wall pinning. Especially defects close the boundary nucleate the sample rapidly. The energy density product decreases with increasing the number of antiphase boundaries. 

Acknowledgment. This work was support by the EU H2020 project Novamag (Grant no 686056) and the Austrian Science Fund FWF (I3288-N36). O.E., H.C. H and E.K.D. also acknowledge support from the Swedish Research Council (VR), the foundation for strategic research (SSF), eSSENCE and STandUPP.


\begin{thebibliography}{22}

\bibitem{Kono}
H. Kono, J. Phys. Soc. Japan {\bf13}, 1444 (1958).

\bibitem{Koch}
A. J. J. Koch, P. Hokkeling, M. G. v. d. Steeg, and K. J. de Vos, J. Appl. Phys. {\bf 30}, 75S, (1960).

\bibitem{Coey}
J. M. D. Coey, J. Phys.: Condens Matter, {\bf 26}, 064221(2014).

\bibitem{Zijlstra}
H. Zijlstra, IEEE Trans. Magnet. {\bf 15}, 1246 (1979).

\bibitem{Bittner}
F. Bittner, L. Schultz, and T. G. Woodcock, Acta Mat. {\bf 101}, 48 (2015).

\bibitem{Houseman}
E. Houseman and J. Jakubovics, J. Mag. Magnet. Mat. {\bf 31-34}, 1007 (1983).

\bibitem{Jakubovics}
J. Jakubovics and T. W. Jolly, Physica B {\bf 86-88B}, 1357 (1977).

\bibitem{Thielsch}
J. Thielsch, F. Bittner, and T. Woodcock, J. Mag. Magnet. Mat. {\bf 426}, 25 (2017).

\bibitem{Bance}
S. Bance, F. Bittner, T. G. Woodckok, L. Schultz, and T. Schrefl,  Acta Mater. {\bf 131}, 48 (2017).

\bibitem{Pnieves}
P. Nieves, S. Arapan, T. Schrefl, and S. Cuesta-L\'{o}pez, Phys. Rev. B \textbf{96}, 224411 (2017).

\bibitem{Eriksson}
 O.Eriksson, A.Bergman, L.Bergqvist, J.Hellsvik, \textit{Atomistic spin-dynamics: foundations and applications}, (Oxford University Press, New York, 2017) 

\bibitem{Liechtenstein}
A. I. Liechtenstein, M. I. Katsnelson, V. P. Antropov, and V. A. Gubanov, J. Magn. Magn. Mater. {\bf67} 65 (1987).

\bibitem{Faulkner}
J. S. Faulkner and G. M. Stocks, Phys. Rev. B {\bf 21}, 3222 (1980).

\bibitem{Rosengaard}
N. M. Rosengaard and B. Johansson, Phys. Rev. B {\bf 55}, 14975 (1997).

\bibitem{Pajda}
M. Pajda, J. Kudrnovsky, I. Turek, V. Drchal, and P. Bruno, Phys. Rev. B {\bf64}, 174402 (2001).


\bibitem{Jaku} J.P. Jakubovics, A.J. Lapworth and T.W. Jolly, J. Appl. Phys. \textbf{49}, 3 (1978).

\bibitem{Land} J. van Landuyt, G. van Tendeloo, J. van den Broek, H. Donkersloot, H. Zijlstra, IEEE Trans. Magnetics 14 (1978) 679.

\bibitem{VASP1}
G. Kresse and J. Furthm{\"u}ller, Phys. Rev. B {\bf54}, 11169 (1996).
 
 \bibitem{VASP2}
 G. Kresse and D. Joubert, Phys. Rev. B {\bf59}, 1758 (1999).

\bibitem{PAW}
J. P. Perdew , K. Burke, and M. Ernzerhof, Phys. Rev. Lett. {\bf77}, 3865 (1996).

\bibitem{Yanar}
C. Yanar,V. Radmilovic,W. A. Soffa, and J. M. Wiezorek, Intermetallics \textbf{9(10-11)}, 949-954 (2001)

\bibitem{Dittrich} R. Dittrich, T. Schrefl, D. Suess, W. Scholz, H. Forster, J. Fidler, J. Appl. Phys. {\bf 93}, 7405, (2003).

\bibitem{Evans}R. F. L. Evans, U. Atxitia, and R. W. Chantrell, Phys. Rev. B {\bf 91}, 144425 (2015).



 \end{thebibliography}
 \end{document}